# The determination of $H_\circ$ by using the TF relation : About particular selection effects

R. Triay[1,2*], S. Rauzy[1,2], and M. Lachièze-Rey[3**]

[1] Centre de Physique Théorique - C.N.R.S., Luminy Case 907, F-13288 Marseille Cedex 9, France
[2] Université de Provence
[3] Service d'Astrophysique, C.E. Saclay, F-91191 Gif sur Yvette Cedex, France

**Abstract.** This paper completes the statistical modeling of the Hubble flow when a Tully-Fisher type relation is used for estimating the absolute magnitude $M \approx a\,p + b$ from a line width distance indicator $p$. Our investigation is performed with the aim of providing us with a full understanding of statistical biases due to selection effects in observation, regardless of peculiar velocities of galaxies. We show that unbiased $H_\circ$-statistics can be obtained by means of the maximum likelihood method as long as the statistical model can be defined. We focus on the statistical models related to the Direct, resp. Inverse, Tully-Fisher relation, when selection effects on distance, resp. on $p$, are present. It turns out that the use of the Inverse relation should be preferred, according to robustness criteria. The formal results are ensured by simulations with samples which are randomly generated according to usual characteristics.

## 1. Introduction

The statistical biases that affect the analysis of cosmic velocity fields, when a Tully-Fisher (TF) relation is used in the distance estimate of galaxies (see Tully & Fisher 1977, Faber & Jackson 1976) is an important problem in observational cosmology (see e.g., Teerikorpi 1975, Schechter 1980). At present, among different approaches, there is a general agreement on the usual bias definition, which says that a *statistic* is biased if its *expected value* does not correspond to the model parameter for which it has been made up (see e.g., Teerikorpi 1994, Strauss & Willick 1994). Such a statement shows clearly the necessity of defining a *statistical model* (e.g., for checking whether a statistic is biased and for calculating the related correction, see e.g., Triay 1993, Willick 1994). With this in mind, we understand that unbiased statistics can be obtained as long as the *probability density* (*pd*) describing the data can be defined. Moreover, nothing prevents us to use solely the *maximum likelihood* (ML) technique, which provides us unambiguously with a unique fitting technique, and prevents us from subjective speculations on diagrams, Triay (1994). In the present fields of interest, the problem of biases is related to the question of whether the selection effects in observation are indeed described by the statistical model. At first glance, there is however an additional difficulty, which is due to the not yet solved problem of the choice between the frequentist and the Bayesian approaches, see Hendry et al. (1994). In this paper, we do not address this question and we limit ourselves to complete the Bayesian approach developed in Triay et al. (1994) investigation (herein TLR), which consists on the statistical modeling of the Hubble flow, regardless of peculiar velocities of galaxies. The aim of such an investigation is to provide us with a full understanding of statistical biases due to selection effects in observation, when the Direct or the Inverse TF models (herein, DTF and ITF) are used (see e.g., Teerikorpi 1982,1984,1987, Bottinelli et al.1985,1986a,1986b,1988a,1988b, Tully 1988a,1988b). A consensus can be found by arguing on the statistical model, instead of the technique of fitting, which shows that the estimates of galaxies distances and $H_0$ are not model dependent, contrarily to calibration parameters of the TF relation (see TLR, Rauzy 1994, Rauzy&Triay 1995). A sensible choice of the model has to be motivated solely by reasons of *robustness* of statistics, which depends on selection effects in observation. Section 3 gives the (unbiased) $H_0$-statistic within the ITF-model when selection effects on the line width distance estimator are present, and Sect. 4 for selection effects on distance (or recession velocity) within the DTF-model (ITF-model with

*Send offprint requests to*: R. Triay
\* the European Cosmological Network
\*\* the European Cosmological Network

tion effects on line width distance estimator are treated in TLR). In order to have a visual support for our theoretical approach and to estimate the magnitude of biases, we perform numerical simulations in Sec. 5. The mathematical framework is specified in Sec. 2. Notations and useful formulas are given in Appendix A, these features are addressed throughout the text by means of symbol "*Def.*".

## 2. The basic model

The difference between the ITF or the DTF relations for obtaining the absolute magnitude of sources $M \approx a\,p + b$, from a log line width distance estimator $p$, interprets essentially within a framework of the statistical modeling of data. The first step of such a process is to enumerate the variables involved in the calculation, which are : – the absolute magnitude $M$, – the line width distance estimator $p$, – the distance modulus $\mu = 25 + 5\log r$ ($r$ is the distance from the observer in Mpc), – and a similarly defined quantity $\eta = 25 + 5\log v$, which accounts for the recession velocity $v$. If the peculiar velocities of sources are neglected then the Hubble law can be written as follows

$$\eta = \mu + \mathcal{H}, \qquad (1)$$

where

$$\mathcal{H} = 5\log H_\circ. \qquad (2)$$

The apparent magnitude is given by,

$$m = M + \mu. \qquad (3)$$

Finally, for clearness in reading, we use the following variables

$$x = m - \eta, \qquad (4)$$
$$y = a.p + b + \eta - m. \qquad (5)$$

The analysis of the Hubble flow involves two steps :

1. *The calibration of the TF-relation* $M \approx a.p + b$. The statistics of model parameters $a$ and $b$ are written in terms of observable quantities $\{p_k, M_k = m_k - \mu_k\}_{k=1, N_1}$.
2. *The determination of the Hubble constant.* The $H_\circ$-statistics are written in terms of observable quantities $\{x_k, y_k\}_{k=1, N_2}$.

The second step is to write the *pd* describing the distribution of above variables (according to working hypotheses) which characterizes the statistical model. Let us write

$$dP_{\text{obs}} = \frac{\phi}{P_{\text{th}}(\phi)}\,dP_{\text{th}}, \qquad (6)$$

where $\phi$ is a function written in terms of observable quantities which describes the selection effects in observation (it is called *selection function*), $dP_{\text{th}}$ is the *pd* describing $P_{\text{th}}(\phi)$ is the normalization factor, see (*Def.*3). For defining the *pd* $dP_{\text{th}}$, one assumes working hypotheses on the distribution of the intrinsic quantities $M$, $\mu$ and $p$. If no luminosity evolutionary effect of sources is present then we can write

$$dP_{\text{th}} = F(M, p)dM\,dp\,\kappa(\mu)d\mu, \qquad (7)$$

where $\kappa(\mu)$ accounts for the distribution of galaxies in space and $F(M, p)$ for the $M$–$p$ distribution. The difference between the ITF and the DTF models lies on describing this *pdf*, one has

$$F(M, p)dM\,dp \approx f_\xi(\xi; \xi_\circ, \sigma_\xi)d\xi\,g(\zeta; 0, \sigma_\zeta)d\zeta, \qquad (8)$$

where $\xi$ depend on the choice of the model which "mimics" the TF-relation :

$$\xi = \begin{cases} M & \text{in the ITF model} \\ p & \text{in the DTF model.} \end{cases} \qquad (9)$$

the parameter $\xi_\circ$, resp. $\sigma_\xi$, denotes the mean, resp. the standard deviation, and the random variable $\zeta$ describes the scatter of the TF-relation, it has necessarily a zero mean, $\sigma_\zeta$ denotes the related standard deviation. We assume :

- $h_0$) a *linear* TF-relation,
$$\zeta = a.p + b - M, \qquad (10)$$
- $h_1$) a Gaussian $\zeta$-distribution
$$g(\zeta; 0, \sigma_\zeta) = g_{\text{G}}(\zeta; 0, \sigma_\zeta). \qquad (11)$$

The description of the $M$–$p$ distribution will be completed in the next sections. About the spatial distribution of sources, we assume that it is given by

- $h_2$) a power law
$$\kappa(\mu) \propto \exp(\beta\mu), \qquad (12)$$

where $\beta = 3\ln 10/5$ accounts for a uniform distribution. A part of selection effects is described in terms of apparent magnitude, (i.e., $\phi \propto \phi_m$) and we assume :

- $h_3$) a cutoff at a given limiting magnitude $m_{\text{lim}}$,
$$\phi_m(m) = \theta(m_{\text{lim}} - m), \qquad (13)$$

where $\theta$ denotes the Heaveside distribution function.

## 3. $p$-limited samples in the ITF model

The ITF-model is specified by the luminosity distribution function, we assume that :

- $h_4$) it is Gaussian
$$f_M(M; M_\circ, \sigma_M) = g_{\text{G}}(M; M_\circ, \sigma_M). \qquad (14)$$

The *pd* describing the data reads

$$dP_{\text{obs}} = \frac{\phi_m(m)\phi_p(p)}{P_{\text{th}}(\phi_m\phi_p)}\,dP_{\text{th}}, \qquad (15)$$

where $\phi_m$ and $dP_{\text{th}}$ are given by Eq. (7-14), $\phi_p(p)$ accounts for $p$-selection effects, and it works as :

$$\phi_p(p) = \theta(p - p_{\min})\theta(p_{\max} - p). \tag{16}$$

The normalization factor $P_{\text{th}}(\phi_m \phi_p)$, see ($Def$.3), reads

$$\begin{aligned}P_{\text{th}}(\phi_m \phi_p) &= \int \theta(p - p_{\min})\theta(p_{\max} - p) a\, dp \\ &\quad g_{\text{G}}(M; M_\circ, \sigma_M) g_{\text{G}}(M; a.p + b, \sigma_\zeta) dM \\ &\quad \theta(m_{\lim} - M - \mu) \exp(\beta\mu) d\mu, \end{aligned} \tag{17}$$

provided the working hypotheses ($h_0$–$h_5$). The inner integration over $\mu$ provides us with the term $\beta^{-1} e^{\beta m_{\lim}} e^{-\beta M}$, and according to ($Def$.2c), we have

$$\begin{aligned}e^{-\beta M} g_{\text{G}}(M; M_\circ, \sigma_M) &= e^{-\beta(M_\circ - \frac{1}{2}\beta\sigma_M^2)} \\ &\quad \times g_{\text{G}}(M; M_\circ - \beta\sigma_M^2, \sigma_M).\end{aligned} \tag{18}$$

Hence the integration over $M$ gives

$$\int g_{\text{G}}(M; M_\circ - \beta\sigma_M^2, \sigma_M) g_{\text{G}}(M; a.p + b, \sigma_\zeta) dM \tag{19}$$

$$= g_{\text{G}}(a.p + b; M_\circ - \beta\sigma_M^2, \sigma^{\text{I}}), \tag{20}$$

where

$$\sigma^{\text{I}} = \sqrt{\sigma_\zeta^2 + \sigma_M^2}, \tag{21}$$

see ($Def$.2d). Finally, we obtain

$$P_{\text{th}}(\phi_m \phi_p) = \mathcal{A}_p P_{\text{th}}(\phi_m), \tag{22}$$

where

$$P_{\text{th}}(\phi_m) = \frac{1}{\beta} \exp\beta\left(m_{\lim} - M_\circ + \frac{\beta}{2}\sigma_M^2\right) \tag{23}$$

does not depend on $a$, $b$ and $\sigma_\zeta$, and

$$\mathcal{A}_p(a, b, \sigma_\zeta) = \omega_p(p_{\max}) - \omega_p(p_{\min}), \tag{24}$$

with

$$\omega_p(p) = \mathcal{N}(v(p)), \tag{25}$$

$$v(p) = \frac{1}{\sigma^{\text{I}}}\left(a\,p + b - M_\circ + \beta\sigma_M^2\right), \tag{26}$$

where $\mathcal{N}$ denotes the erfc function, see ($Def$.1c).

3.1. Calibration statistics

According to TLR, if $p$-selection effects are not present in the observations then the model parameters $a$, $b$ and $\sigma_\zeta$ can be measured as follows

$$a^{\text{ITF}} = \frac{\Sigma_1(M)^2}{\text{Cov}_1(p, M)}, \tag{27}$$

$$b^{\text{ITF}} = \langle M \rangle_1 - a^{\text{ITF}} \langle p \rangle_1, \tag{28}$$

$$\sigma_\zeta^{\text{ITF}} = \Sigma_1(M) \sqrt{\frac{1}{\rho_1^2(p, M)} - 1}. \tag{29}$$

$\phi_m$, $f_M$ and $\kappa$. In the other hand, if $p$-selection effects are present then these functions have to be specified for obtaining calibration statistics. The only reason is that the normalization factor depends on model parameters, through $\mathcal{A}_p(a, b, \sigma_\zeta)$ which reads in terms of $\phi_m$, $f_M$ and $\kappa$, see Eq. (22). According to working hypotheses ($h_0$–$h_5$), the efficient part of the likelihood function ($lf$), see ($Def$.4), is given by

$$\mathcal{L}_1^{\text{I}}(a, b, \sigma_\zeta) = -\ln \mathcal{A}_p(a, b, \sigma_\zeta) + \mathcal{L}_1^{\text{ITF}}(a, b, \sigma_\zeta), \tag{30}$$

where

$$\mathcal{L}_1^{\text{ITF}} = \ln a - \ln \sigma_\zeta - \frac{1}{N_1} \sum_{k=1}^{N_1} \frac{(a.p_k + b - M_k)^2}{2\sigma_\zeta^2} \tag{31}$$

is the likelihood function when no $p$-selection effect is present. For convenience in notation, let us define the functions

$$\vartheta_\alpha(a, b, \sigma_\zeta) = \frac{\partial_\alpha \mathcal{A}_p}{\mathcal{A}_p}, \tag{32}$$

with $\alpha = a, b, \sigma_\zeta^2$. Hence, the ML equations are given by

$$a \langle p(ap + b - M) \rangle_1 = \sigma_\zeta^2 (1 - a\vartheta_a), \tag{33}$$

$$\langle ap + b - M \rangle_1 = -\sigma_\zeta^2 \vartheta_b, \tag{34}$$

$$\langle (ap + b - M)^2 \rangle_1 = \sigma_\zeta^2 \left(1 + 2\sigma_\zeta^2 \vartheta_{\sigma_\zeta^2}\right), \tag{35}$$

see ($Def$.5). According to Eq. (24-26), the generic form for these terms reads

$$\partial \mathcal{A}_p(a, b, \sigma_\zeta) = \partial \omega_p(p_{\max}) - \partial \omega_p(p_{\min}) \tag{36}$$

where

$$\partial_a \omega_p(p) = \frac{p\, g_{\text{N}}(v(p))}{\sigma^{\text{I}}}, \tag{37}$$

$$\partial_b \omega_p(p) = \frac{g_{\text{N}}(v(p))}{\sigma^{\text{I}}}, \tag{38}$$

$$\partial_{\sigma_\zeta^2} \omega_p(p) = -\frac{v(p)\, g_{\text{N}}(v(p))}{2(\sigma^{\text{I}})^2}. \tag{39}$$

Equation (34) gives straightforwardly the $b$-statistic. The $a$-statistic is obtained by substituting $b$ into Eq. (33,35), according to Eq. (34). By arranging the terms in descending order of magnitude, we obtain[1]

$$a = a^{\text{ITF}} \left\{ 1 + \gamma_{\text{I}}^2 \left( a \left(\langle p \rangle_1 \vartheta_b - \vartheta_a\right) + \sigma_\zeta^2 \left(\vartheta_b^2 - 2\vartheta_{\sigma_\zeta^2}\right) \right) \right\}, \tag{40}$$

where $a^{\text{ITF}}$ is given in Eq. (27), and

$$\gamma_{\text{I}} = \frac{\sigma_\zeta}{\Sigma_1(M)}, \tag{41}$$

---

[1] A trick, which prevents us from cumbersome calculation: the following equality $\langle ap\,(a(p - \langle p \rangle_1) - (M - \langle M \rangle_1)) \rangle_1 = \langle a\,(p - \langle p \rangle_1)\,(a\,(p - \langle p \rangle_1) - (M - \langle M \rangle_1)) \rangle_1$

$$\langle (ap+b-M)^2 \rangle_1 = \langle ap(ap+b-M) \rangle_1 + b\langle (ap+b-M) \rangle_1 \\ - \langle M(ap+b-M) \rangle_1. \quad (42)$$

The first two terms are substituted according to Eq. (33,34). The third one reads $a\mathrm{Cov}(p,M) - \Sigma_1^2(M) - \sigma_\zeta^2 \langle M \rangle_1 \vartheta_b$. Hence, we obtain an expression in terms of $\sigma_\zeta^2$ and $\sigma_\zeta^4$, and the $\sigma_\zeta^2$-statistic interprets as the root of a second order polynomial. Finally, the ML statistics are given by

$$a^{\mathrm{I}} = a^{\mathrm{ITF}} \times \frac{1+\gamma_{\mathrm{I}}^2 \Psi_1^{\mathrm{I}}}{1+\gamma_{\mathrm{I}}^2 \Psi_2^{\mathrm{I}}}, \quad (43)$$

$$b^{\mathrm{I}} = \langle M \rangle_1 - a^{\mathrm{I}} \langle p \rangle_1 - \left(\sigma_\zeta^{\mathrm{I}}\right)^2 \vartheta_b, \quad (44)$$

$$\left(\sigma_\zeta^{\mathrm{I}}\right)^2 = \frac{1}{4\vartheta_{\sigma_\zeta^2}} \left( \sqrt{1 + 8\vartheta_{\sigma_\zeta^2} \langle (a^{\mathrm{I}} p + b^{\mathrm{I}} - M)^2 \rangle_1} - 1 \right), \quad (45)$$

where the $\Psi_i^{\mathrm{I}}$ are functions defined by

$$\Psi_1^{\mathrm{I}}(a,b,\sigma_\zeta) = \sigma_\zeta^2 \left( \vartheta_b^2 - 2\vartheta_{\sigma_\zeta^2} \right), \quad (46)$$

$$\Psi_2^{\mathrm{I}}(a,b,\sigma_\zeta) = a^{\mathrm{ITF}} \left( \vartheta_a - \langle p \rangle_1 \vartheta_b \right). \quad (47)$$

In practice, the estimates of parameters $a \approx a^{\mathrm{I}}$, $b \approx b^{\mathrm{I}}$ and $\sigma_\zeta \approx \sigma_\zeta^{\mathrm{I}}$ are obtained from Eq. (43-45) by Newton iterative method. The derivation of pdf related to calibration statistics is cumbersome, and useless because of the small number of sources that are used in general. The statistical behavior for these estimates is investigated by means of simulations, see TLR.

### 3.2. Determination of $H_\circ$

The pd given in Eq. (6) reads in terms of observable quantities $x$, $y$ and $\eta$, see Eq. (1,4,5), as follows

$$dP_{\mathrm{obs}} = \frac{\phi_m(x+\eta)}{P_{\mathrm{th}}(\phi_m \phi_p)} f_M(x; M_\circ - \mathcal{H}, \sigma_M) \kappa(\eta - \mathcal{H}) dx d\eta \\ \times \phi_p(\frac{x+y-b}{a}) g_G(y; \mathcal{H}, \sigma_\zeta) dy. \quad (48)$$

Let us emphasize that if p-selection effects are not present then this pd reads as a product of two independent pds, which provided us with a robust $H_\circ$-statistic

$$\mathcal{H}^{\mathrm{ITF}^\star} = \langle y \rangle_2, \quad (49)$$

see TLR. Thus this statistic becomes biased because of the extra term $\phi_p$ in Eq. (48). In the other hand, the ML statistic defined by

$$\mathcal{H}^{\mathrm{ITF}} = \frac{\mathcal{H}^{\mathrm{ITF}^\star} + \gamma^2 \mathcal{H}^{\mathrm{C}}}{1+\gamma^2}, \quad (50)$$

depend on the Hubble constant $H_\circ$ for every functions $\phi_m$ and $\phi_p$[2]. One has

$$\gamma = \frac{\sigma_\zeta}{\sigma_M} \approx \gamma_{\mathrm{I}}, \quad (51)$$

and

$$\mathcal{H}^{\mathrm{C}} = \left( M_\circ - \beta \sigma_M^2 \right) - \langle x \rangle_2. \quad (52)$$

The $H_\circ$-statistic given in Eq. (50) is Gaussian distributed about $\mathcal{H}$ with a standard deviation given by

$$\sigma_{\mathcal{H}^{\mathrm{ITF}}} = \frac{1}{\sqrt{1+\gamma^2}} \frac{\sigma_\zeta^{\mathrm{I}}}{\sqrt{N_2}}, \quad (53)$$

see Eq. (45).

## 4. $\mu$-limited samples in the DTF model

The DTF-model is specified by the p-distribution function, and we assume that :

- $h_4'$) it is Gaussian

$$f_p(p; p_\circ, \sigma_p) = g_G(p; p_\circ, \sigma_p). \quad (54)$$

The pd describing the data reads

$$dP_{\mathrm{obs}} = \frac{\phi_m(m)\phi_\mu(\mu)}{P_{\mathrm{th}}(\phi_m \phi_\mu)} dP_{\mathrm{th}}, \quad (55)$$

where $\phi_m$ and $dP_{\mathrm{th}}$ are given by Eq. (7-10,13-11,54), $\phi_\mu(\mu)$ accounts for $\mu$-selection effects, and it works as :

- $h_6$) a window distribution function

$$\phi_\mu(\mu) = \theta(\mu - \mu_{\min}) \theta(\mu_{\max} - \mu). \quad (56)$$

Hence, the normalization factor $P_{\mathrm{th}}(\phi_m \phi_\mu)$, see (Def.3), within hypotheses ($h_0$-$h_3$,$h_4'$,$h_6$) reads

$$P_{\mathrm{th}}(\phi_m \phi_\mu) = \int \theta(\mu - \mu_{\min}) \theta(\mu_{\max} - \mu) \\ \theta(m_{\lim} - M - \mu) \exp(\beta \mu) dM d\mu \\ g_G(p; p_\circ, \sigma_p) g_G(a.p+b; M, \sigma_\zeta) dp. \quad (57)$$

The inner integration over $p$ provides us with the term $g_G(M; a.p_\circ + b, \sigma^{\mathrm{D}})$, see (Def.2b,d), where

$$\sigma^{\mathrm{D}} = \sqrt{a^2 \sigma_p^2 + \sigma_\zeta^2}. \quad (58)$$

Hence, the $\mu$-$M$ domain of integration can be separated into two sub-domains : the first one is defined by $M \leq$

---

[2] It becomes obvious by writing

$$P_{\mathrm{th}}(\phi_m \phi_p) = \int \phi_m(x'+\eta') \phi_p(\frac{x'+y'-b}{a}) f_M(x'; M_\circ, \sigma_M) \\ \times \kappa(\eta') dx' d\eta' g_G(y'; 0, \sigma_\zeta) dy',$$

where the dummy variables $x' = x + \mathcal{H}$, $y' = y - \mathcal{H}$ and $\eta' = \eta - \mathcal{H}$.

$m_{\lim}-\mu_{\max} < M \leq m_{\lim}-\mu_{\min}$ and $\mu_{\min} < \mu \leq m_{\lim}-M$. Once the integration over $\mu$ is carried out, the one over $M$ terminates the calculation. Finally, by using the following variables

$$\mu_\circ = m_{\lim} - (a.p_\circ + b), \qquad (59)$$

$$\omega_\mu(\mu) = \mathcal{N}(v_2(\mu)) \exp(v_1(\mu)) - \mathcal{N}(v_3(\mu)), \qquad (60)$$

$$v_1(\mu) = \beta\sigma^{\mathrm{D}} \left( \frac{\mu - \mu_\circ}{\sigma^{\mathrm{D}}} - \beta\frac{\sigma^{\mathrm{D}}}{2} \right), \qquad (61)$$

$$v_2(\mu) = \frac{\mu - \mu_\circ}{\sigma^{\mathrm{D}}}, \qquad (62)$$

$$v_3(\mu) = \frac{\mu - \mu_\circ}{\sigma^{\mathrm{D}}} + \beta\sigma^{\mathrm{D}}, \qquad (63)$$

see ($Def$.1b,c), the normalization factor can be written

$$P_{\mathrm{th}}(\phi_m \phi_\mu) = \mathcal{A}_\mu P_{\mathrm{th}}(\phi_m), \qquad (64)$$

where

$$P_{\mathrm{th}}(\phi_m) = \frac{1}{\beta} \exp \beta \left( \mu_\circ + \frac{\beta}{2} \left( a^2 \sigma_p^2 + \sigma_\zeta^2 \right) \right), \qquad (65)$$

and

$$\mathcal{A}_\mu(a, b, \sigma_\zeta) = \omega_\mu(\mu_{\max}) - \omega_\mu(\mu_{\min}). \qquad (66)$$

### 4.1. Calibration statistics

According to TLR, if $\mu$-selection effects are not present in the observations then the model parameters $a$, $b$ and $\sigma_\zeta$ can be estimated by the following statistics

$$a^{\mathrm{DTF}} = \frac{\mathrm{Cov}_1(p, M)}{(\Sigma_1(p))^2}, \qquad (67)$$

$$b^{\mathrm{DTF}} = \langle M \rangle_1 - a^{\mathrm{DTF}} \langle p \rangle_1 + \beta \left( \sigma_\zeta^{\mathrm{DTF}} \right)^2, \qquad (68)$$

$$\sigma_\zeta^{\mathrm{DTF}} = \Sigma_1(M) \sqrt{1 - \rho_1^2(p, M)}, \qquad (69)$$

provided hypotheses ($h_0$–$h_3$,$h_4'$). When $\mu$-selection effects are present then $\mathcal{A}_\mu$ intervenes in the normalization factor, see Eq.(64). Since, both $P_{\mathrm{th}}(\phi_m)$ and $\mathcal{A}_\mu$ depend on $a$ and $b$, the efficient part of the likelihood function reads

$$\mathcal{L}_1^{\mathrm{D}}(a, b, \sigma_\zeta) = -\ln \mathcal{A}_\mu(a, b, \sigma_\zeta) + \mathcal{L}_1^{\mathrm{DTF}}(a, b, \sigma_\zeta), \qquad (70)$$

where

$$\mathcal{L}_1^{\mathrm{DTF}} = -\ln P_{\mathrm{th}}(\phi_m) - \ln \sigma_\zeta - \frac{1}{N_1} \sum_{k=1}^{N_1} \frac{(a.p_k + b - M_k)^2}{2\sigma_\zeta^2}$$

$$-\ln \sigma_p - \frac{1}{N_1} \sum_{k=1}^{N_1} \frac{(p_k - p_0)^2}{2\sigma_p^2}, \qquad (71)$$

is the likelihood function when no $\mu$-selection effect is present. Hence, the ML equations are given by

$$\langle p(ap + b - M) \rangle_1 = \sigma_\zeta^2 \left\{ \beta(p_\circ - \beta a \sigma_p^2) - \varrho_a \right\}, \qquad (72)$$

$$\langle ap + b - M \rangle_1 = \sigma_\zeta^2 (\beta - \varrho_b), \qquad (73)$$

$$\langle (ap + b - M)^2 \rangle_1 = \sigma_\zeta^2 \left( 1 + \sigma_\zeta^2 (\beta^2 + 2\varrho_{\sigma_\zeta^2}) \right). \qquad (74)$$

$$\langle (p - p_\circ) \rangle_1 = \sigma_p^2 (\beta a + \varrho_{p_\circ}), \qquad (75)$$

$$\langle (p - p_\circ)^2 \rangle_1 = \sigma_p^2 \left\{ 1 + \sigma_p^2 \left( a^2 \beta^2 - 2\varrho_{\sigma_p^2} \right) \right\}, \qquad (76)$$

where (for convenience in notation) the functions

$$\varrho_\alpha(a, b, \sigma_\zeta) = \frac{\partial_\alpha \mathcal{A}_\mu}{\mathcal{A}_\mu}, \quad (\alpha = a, b, \sigma_\zeta^2, p_\circ, \sigma_p^2). \qquad (77)$$

According to Eq.(59-63,66), these last terms are calculated as follows. One has

$$\partial \mathcal{A}_\mu(a, b, \sigma_\zeta) = \partial \omega_\mu(\mu_{\max}) - \partial \omega_\mu(\mu_{\min}) \qquad (78)$$

where

$$\partial \omega_\mu(\mu) = (\partial v_2 g_{\mathrm{N}}(v_2) + \partial v_1 \mathcal{N}(v_2)) \exp(v_1) - \partial v_3 g_{\mathrm{N}}(v_3), (79)$$

with

$$\partial_a v_1(\mu) = \beta(p_\circ - \beta a \sigma_p^2), \qquad (80)$$

$$\partial_a v_2(\mu) = -\frac{1}{\sigma^{\mathrm{D}}} \left( p_\circ - \sigma_p \frac{(\mu - \mu_\circ)}{\sigma_{\mathrm{DTF}}} \frac{a \sigma_p}{\sigma^{\mathrm{D}}} \right), \qquad (81)$$

$$\partial_a v_3(\mu) = \partial_a v_2(\mu) + \beta \sigma_p \frac{a \sigma_p}{\sigma^{\mathrm{D}}}, \qquad (82)$$

$$\partial_b v_1(\mu) = \beta, \qquad (83)$$

$$\partial_b v_2(\mu) = -\frac{1}{\sigma^{\mathrm{D}}}, \qquad (84)$$

$$\partial_b v_3(\mu) = \partial_b v_2(\mu), \qquad (85)$$

$$\partial_{\sigma_\zeta^2} v_1(\mu) = -\frac{\beta^2}{2}, \qquad (86)$$

$$\partial_{\sigma_\zeta^2} v_2(\mu) = \frac{1}{2(\sigma^{\mathrm{D}})^2} \frac{(\mu - \mu_\circ)}{\sigma^{\mathrm{D}}}, \qquad (87)$$

$$\partial_{\sigma_\zeta^2} v_3(\mu) = \partial_{\sigma_\zeta^2} v_2(\mu) + \frac{\beta}{2\sigma^{\mathrm{D}}}, \qquad (88)$$

$$\partial_{p_\circ} v_i(\mu) = a \partial_b v_i(\mu), \qquad (89)$$

$$\partial_{\sigma_p^2} v_i(\mu) = a^2 \partial_{\sigma_\zeta^2} v_i(\mu), \qquad (90)$$

where $i = 1, 2, 3$ for Eq.(89,90). Finally, the ML equations yield the following statistics :

$$a^{\mathrm{D}} = a^{\mathrm{DTF}} \times \frac{1 + \gamma_{\mathrm{D}}^2 \Psi_1^{\mathrm{D}}(a, b, \sigma_\zeta)}{1 - \gamma_{\mathrm{D}}^2 \Psi_2^{\mathrm{D}}(a, b, \sigma_\zeta)}, \qquad (91)$$

$$b^{\mathrm{D}} = \langle M \rangle_1 - a^{\mathrm{D}} \langle p \rangle_1 + \left( \sigma_\zeta^{\mathrm{D}} \right)^2 (\beta - \varrho_b), \qquad (92)$$

$$\left( \sigma_\zeta^{\mathrm{D}} \right)^2 = \frac{1}{\Psi_3^{\mathrm{D}}} \left( \sqrt{1 + 2\Psi_3^{\mathrm{D}} \langle (a^{\mathrm{D}} p + b^{\mathrm{D}} - M)^2 \rangle_1} - 1 \right) \quad (93)$$

and (if necessary)

$$p_\circ^{\mathrm{D}} = \langle p \rangle_1 + a^{\mathrm{D}} \left( \sigma_p^{\mathrm{D}} \right)^2 (\beta - \varrho_b), \qquad (94)$$

$$\left( \sigma_p^{\mathrm{D}} \right)^2 = \frac{1}{\Psi_3^{\mathrm{D}}} \left( \sqrt{1 + 2\Psi_3^{\mathrm{D}} \langle (p - p_\circ)^2 \rangle_1} - 1 \right), \qquad (95)$$

defined by

$$\gamma_D = \frac{\sigma_\zeta}{\sqrt{-\text{Cov}_1(p, M)}}, \quad (96)$$

provided $\text{Cov}_1(p, M) < 0$, and the functions $\Psi_i^D$ read

$$\Psi_1^D(a, b, \sigma_\zeta) = (\varrho_a - \langle p \rangle_1 \varrho_b) \quad (97)$$
$$\Psi_2^D(a, b, \sigma_\zeta) = a^{\text{ITF}} \varrho_b \sigma_p^2, \quad (98)$$
$$\Psi_3^D(a, b, \sigma_\zeta) = 2\left(\beta^2 + 2\varrho_{\sigma_\zeta^2}\right). \quad (99)$$

Similarly as previous section, the estimates of parameters $a \approx a^D$, $b \approx b^D$ and $\sigma_\zeta \approx \sigma_\zeta^D$ are obtained from Eq. (91-95) by Newton iterative method.

### 4.2. Determination of $H_o$

The terminology "distance limited sample" is meaningless in this step, because $\mu$ is not an observable, and thus one has to account for selection effects on recession velocities. Let us assume that the $\eta$-selection effects are described by:

- $h_7$) a window distribution function

$$\phi_\eta(\eta) = \theta(\eta - \eta_{\min})\theta(\eta_{\max} - \eta). \quad (100)$$

The pd, given in Eq. (6), reads in terms of observable quantities $x$, $y$ and $\eta$, see Eq. (1,4,5), as follows

$$dP_{\text{obs}} = \frac{\phi_m(x+\eta)\phi_\eta(\eta)}{P_{\text{th}}(\phi_m \phi_\eta)} \frac{1}{a} f_p\left(\frac{x+y-b}{a}; p_o, \sigma_p\right)$$
$$\kappa(\eta - \mathcal{H}) dx d\eta \times g_G(y; \mathcal{H}, \sigma_\zeta) dy. \quad (101)$$

where $\phi_m$, $f_p$ and $\kappa$ are given by Eq(13,54,12). The calculation of the normalization factor $P_{\text{th}}(\phi_m \phi_\eta)$ is obvious by using the results of previous section,

$$P_{\text{th}}(\phi_m \phi_\eta) = \mathcal{A}_\eta P_{\text{th}}(\phi_m), \quad (102)$$

where $P_{\text{th}}(\phi_m)$ is given by Eq.(65), and

$$\mathcal{A}_\eta(\mathcal{H}) = \omega_\mu(\eta_{\max} - \mathcal{H}) - \omega_\mu(\eta_{\min} - \mathcal{H}). \quad (103)$$

Therefore, according to TLR, the lf reads

$$\mathcal{L}_2^{\text{DTF}}(\mathcal{H}) = -\ln \mathcal{A}_\eta - \frac{1}{N_2} \sum_{k=1}^{N_2} \frac{(y_k - \mathcal{H})^2}{2\sigma_\zeta^2} - \beta\mathcal{H}, \quad (104)$$

which shows that the statistic

$$\mathcal{H}^{\text{DTF}} = \langle y \rangle_2 - \beta \sigma_\zeta^2, \quad (105)$$

must be corrected as follows

$$\mathcal{H}^D = \mathcal{H}^{\text{DTF}} - \sigma_\zeta^2 \frac{\partial_\mathcal{H} \mathcal{A}_\eta}{\mathcal{A}_\eta}(\mathcal{H}). \quad (106)$$

term depends on $\mathcal{H}$, which forces us to obtain the $\mathcal{H}$-estimate by iterative method. Let us mention that we have

$$\partial_\mathcal{H} \mathcal{A}_\eta = \partial_b \omega_\mu(\eta_{\min} - \mathcal{H}) - \partial_b \omega_\mu(\eta_{\max} - \mathcal{H}), \quad (107)$$

see Eq. (79,83-85). According to TLR, the statistic given in Eq. (106) is Gaussian distributed about $\mathcal{H}$ with a standard deviation

$$\sigma_{\mathcal{H}^D} = \frac{\sigma_\zeta^D}{\sqrt{N_2}}. \quad (108)$$

see Eq. (69).

### 5. Magnitude of biases – Simulations

In this section, we investigate the biases (i.e., correction terms) of statistics defined in TLR when $p$-selection effects, resp. $\mu$-selection or $\eta$-selection effects, are present. Namely, for the calibration step, let us define

$$\delta a^I = a^I - a^{\text{ITF}}, \quad \delta b^I = b^I - b^{\text{ITF}},$$
$$\delta a^D = a^D - a^{\text{DTF}}, \quad \delta b^D = b^D - b^{\text{DTF}}, \quad (109)$$

see Eq. (27,28,43,44,67,67,91,92) and the biases of $H_o$-statistics: $\delta \mathcal{H}^I$ is the bias of the ITF statistic given in Eq. (50), when $a^{\text{ITF}}$ and $b^{\text{ITF}}$ are used instead of $a^I$ and $b^I$; $\delta \mathcal{H}^D$ is the bias of the DTF statistic given in Eq. (105). In order to have a visual support for our theoretical approach and to estimate the magnitude of biases we use numerical simulations. The random samples are performed according to previous hypothesis. We assume a Hubble's constant of $H_o = 100$ Mpc/km.s$^{-1}$, a cut-off at apparent magnitude of $m_{\lim} = 12$, and additional selection effects that depend on the TF model. The sample-size is chosen sufficiently large for minimizing the effect of statistical fluctuations in the analysis of bias.

It turns out that these simulations validate the statistics defined in previous sections, and in particular the efficiency of the Newton iterative method, since we obtained the same values of TF parameters and of $H_o$ that were assumed for the simulations.

### 5.1. Simulation of p-limited samples in the ITF model

For generating $p$-limited samples, we assume a Gaussian luminosity distribution with $M_o = -19$ and $\sigma_M = 1.5$; a TF-diagram defined by a slope of $a^{\text{ITF}} = -6$ and a zero point $b^{\text{ITF}} = -7$, with a scatter of $\sigma_\zeta^{\text{ITF}} = 0.5$; the $p$-selection effects described by a window distribution function with $p_{\min} = 2$ and $p_{\max} = 3$.

The magnitude of biases read

$$\delta a^I = \gamma_I^2 a^{\text{ITF}} \frac{\Psi_1^I - \Psi_2^I}{1 + \gamma_{\text{ITF}}^2 \Psi_2^I}, \quad (110)$$

$$\delta b^I = -\delta a^I \langle p \rangle_1 - \vartheta_b \sigma_\zeta^2, \quad (111)$$

$$\delta \mathcal{H}^I = \frac{\mathcal{C}_p}{1 + \gamma^2} \left(\delta a^I \Sigma_1(p) - \vartheta_b \left(\sigma_\zeta^I\right)^2\right), \quad (112)$$

$$\mathcal{C}_p = \frac{\langle p \rangle_1 - \langle p \rangle_2}{\Sigma_1(p)}. \tag{113}$$

It is interesting to note in Eq. (112) that the bias of the ITF statistics is canceled if the $\mathcal{C}_p$-*criteria* (i.e., $\mathcal{C}_p = 0$) is fulfilled (see TLR) when the calibration ITF statistics provide us with biased estimates. The simulations provided us with the following numerical values

$$\delta a^{\mathrm{I}} \approx 0.21, \quad \delta b^{\mathrm{I}} \approx 0.45, \quad \delta \mathcal{H}^{\mathrm{I}} \approx 0.01, \tag{114}$$

which shows that the bias in estimating the Hubble constant is $\sim 0.5$ percent.

### 5.2. Simulation of $\mu$-limited samples in the DTF model

For $\mu$-limited samples, we assume a Gaussian $p$-distribution with $p_\mathrm{o} = 2.5$ and $\sigma_p = 0.25$; a TF-diagram defined by $a^{\mathrm{DTF}} = -6$, $b^{\mathrm{DTF}} = -7$ and $\sigma_\zeta^{\mathrm{DTF}} = 0.5$; for the calibration sample the $\mu$-selection effects described by a window distribution function with $\mu_{\min} = 30$ and $\mu_{\max} = 32$; for the sample used for estimating $H_\mathrm{o}$, the $\eta$-selection described by a window distribution function with $\eta_{\min} = 40$ and $\eta_{\max} = 42$.

The magnitude of biases reads

$$\delta a^{\mathrm{D}} = \gamma_{\mathrm{D}}^2 \, a^{\mathrm{DTF}} \frac{\Psi_1^{\mathrm{D}} + \Psi_2^{\mathrm{D}}}{1 - \gamma_{\mathrm{D}}^2 \Psi_2^{\mathrm{D}}}, \tag{115}$$

$$\delta b^{\mathrm{D}} = -\delta a^{\mathrm{D}} \langle p \rangle_1 + \beta \delta \sigma_\zeta^{\mathrm{D}} \left( 2\sigma_\zeta^{\mathrm{D}} - \delta \sigma_\zeta^{\mathrm{D}} \right) - \varrho_b \left( \sigma_\zeta^{\mathrm{D}} \right)^2, \tag{116}$$

$$\delta \mathcal{H}^{\mathrm{D}} = \mathcal{C}_p \delta a^{\mathrm{D}} \Sigma_1(p) + \beta \delta \sigma_\zeta^{\mathrm{D}} \left( 2\sigma_\zeta^{\mathrm{D}} - \delta \sigma_\zeta^{\mathrm{D}} \right) - \varrho_b \left( \sigma_\zeta^{\mathrm{D}} \right)^2, \tag{117}$$

where $\mathcal{C}_p$ is given in Eq. (113). The simulations provided us with the following numerical values

$$\delta a^{\mathrm{D}} \approx 0.15, \quad \delta b^{\mathrm{D}} \approx 0.01, \quad \delta \mathcal{H}^{\mathrm{D}} \approx 0.31, \tag{118}$$

which shows that the bias in estimating the Hubble constant is $\sim 14$ percent.

## 6. Conclusion

This statistical model of the Hubble flow, when a Tully-Fisher (TF) type relation is used for estimating the absolute magnitude $M \approx a\,p + b$ from a line width distance indicator $p$, is in agreement and completes the previous results obtained in Triay et al. (1994) (TLR). Namely, the "Direct" and the "Inverse" TF methods identify to maximum likelihood approaches corresponding to different models of the TF diagram. Hence, coherent estimates of model independent parameters, such as $H_\mathrm{o}$ and the galaxies distances, are obtained from (model dependent) unbiased statistics as long as the same model is used in the calibration step. The choice of the model should be motivated solely by criteria of accuracy and robustness of robust the statistics), which depends on selection effects in observation. For the calibration of the TF relation, if $p$-selection effects are not present then the most attractive approach for the determination of $H_\mathrm{o}$ corresponds to the ITF model, because not information on the data sample is required (about the spatial distribution of sources, the luminosity distribution function and the sample completeness on apparent luminosity and distance) for obtaining the calibration statistics. Otherwise, in addition of these characteristics, one has to be able to describe all the selection effects by specifying the form of the corresponding selection functions. In the DTF model, the data characteristics must be wholly specified for the derivation of unbiased calibration statistics.

The present investigation provides us with unbiased statistics for the ITF model with $p$-selection effects and for the DTF model with selection effects on distance. While the calculation for obtaining unbiased statistics within both models which accounts simultaneously for these two selection effects is straightforward, it is not performed in our analysis. The formulas are derived by assuming usual working hypotheses : a Gaussian luminosity distribution function, a power law spatial distribution of sources, completeness up to a limiting apparent magnitude and window selection functions for describing the $p$-selection effects and selection effects on distance. These statistics has been checked successfully by numerical simulations by using random samples with usual characteristics. The biases due to these selection effects when using the previous statistics (given in TLR) are investigated, and their expected magnitudes are provided by the simulations. It turns out that the bias in estimating the Hubble constant within the ITF model can be removed by using the $\mathcal{C}_p$-*criteria*. It is interesting to note that this bias is much weaker (0.5 percent) in the ITF model than the one (14 percent) in the DTF model, while the order of magnitude of biases in the calibration parameters are larger than those in the DTF model.

Finally, let us emphasize that such a statistical framework can be used for obtaining likely distance estimates of galaxies in the usual sense[4], see Triay (1993).

*Acknowledgements.* Financial support from *GdR Cosmologie* has been appreciated.

## A. Notations and useful formulas

The following features are addressed throughout the text by using the symbol "*Def.*".

*Def.*1 The *probability density (pd)* of a random variable $x$ reads $dP(x) = f(x)dx$, where $f(x)$ represents the *pd function*

---

[3] It is interesting to mention that these characteristics are almost antagonistic.

[4] While it is evident that a more sensible approach requires a statistical modeling of the underlying dynamics of cosmic velocity fields.

exhibit the *model parameters* involved in the statistical model, as the mean $x_0$ and the standard deviation $\sigma$, by writing $f(x; x_0, \sigma)$.
  (a) $g_G(x; x_0, \sigma) = (\sigma\sqrt{2\pi})^{-1} \exp-\left((x-x_0)^2/(2\sigma^2)\right)$ is a Gaussian *pdf*.
  (b) $g_N(x) = g_G(x; 0, 1)$ is a Normal *pdf*.
  (c) $\mathcal{N}(x) = \int_{-\infty}^{x} g_N(t) dt$ is the cumulative Normal *pdf*.

*Def.*2 Let $f$ be a *pdf*, and $\lambda$ be a scalar value, in most of calculations, we use the following properties :
  (a) $f(x + \lambda; x_0, \sigma) = f(x; x_0 - \lambda, \sigma)$;
  (b) $f(\lambda x; x_0, \sigma) = \lambda^{-1} f(x; \frac{x_0}{\lambda}, \frac{\sigma}{\lambda})$;
  (c) $\exp(\lambda x) g_G(x; x_0, \sigma) = \exp\left(\lambda(x_0 + \lambda\frac{\sigma^2}{2})\right) g_G(x; x_0 + \lambda\sigma^2, \sigma)$.
  (d) $g_G(x; x_1, \sigma_1) g_G(x; x_2, \sigma_2) = g_G(x; x_0, \sigma_0) g_G(x_1; x_2, \acute{\sigma})$, where $\acute{\sigma} = \sqrt{\sigma_1^2 + \sigma_2^2}$, $x_0$ and $\sigma_0$ are defined as follows $\sigma_0^{-2} = \sigma_1^{-2} + \sigma_2^{-2}$ and $x_0 \sigma_0^{-2} = x_1 \sigma_1^{-2} + x_2 \sigma_2^{-2}$.

*Def.*3 $P(h) = \int h(x) dP(x)$ denotes the expected value of the function $h(x)$.

*Def.*4 The *pd* of a sample data $\{\mathcal{G}_k\}_{k=1,N}$, which consists of $N$ independently selected objects $\mathcal{G}_k = (x_k, y_k, \ldots)$, is given by $\prod_{k=1}^{N} dP(\mathcal{G}_k)$.
  (a) Its *pdf*, written in terms of *observable quantities* $x_k, y_k, \ldots$, but regarded as a function of model parameters, provides us with the *likelihood function*.
  (b) (*The* ML *method*.) The statistics of model parameters are obtained by maximizing the likelihood function, or (equivalently) its natural logarithm. Actually, the terms which do not contribute to the determination of parameters are removed from it. Such a quantity is called the *efficient part* of likelihood function, herein simply denoted *lf*.

*Def.*5 We have the usual definitions :
  (a) $\langle x \rangle = \sum_{k=1}^{N} x_k / N$ is the sample average,
  (b) $\text{Cov}(x, y) = \sum_{k=1}^{N} (x_k - \langle x \rangle)(y_k - \langle y \rangle)/(N-1)$ is the covariance,
  (c) $\Sigma(x) = \sqrt{\text{Cov}(x, x)}$ is the standard deviation,
  (d) $\rho(x, y) = \text{Cov}(x, y)/(\Sigma(x)\Sigma(y))$ is the correlation coefficient.
  (e) These symbols are written as follows $N_s$, $\langle . \rangle_s$, $\text{Cov}_s(.,.)$, $\Sigma_s(.)$ and $\rho_s(.,.)$ in order to distinguish between the calibration step ($s=1$), and the $H_o$-determination ($s=2$).

*Def.*6 The accuracy of an estimator is formally defined as the reciprocal of its variance (The smaller the dispersion, the greater the precision.).